\documentclass[reprint,twocolumn,prb,amsmath,amssymb,aps]{revtex4-2} 

\usepackage{graphicx}  
\usepackage{dcolumn}   
\usepackage{bm}        
\usepackage{hyperref}  
\hypersetup{colorlinks=true, linkcolor=blue!95!black!85!yellow, citecolor=blue!95!black!85!yellow, urlcolor=blue!95!black!85!yellow} 
\usepackage{physics}   
\usepackage{newtxtext} 
\usepackage[cmintegrals]{newtxmath} 
\usepackage{booktabs}  
\usepackage{multirow}  
\usepackage{xcolor}    
\usepackage{float}     

\usepackage{tikz}
\newcommand{\orcidicon}{%
    \begin{tikzpicture}
        \draw[lime, fill=lime] (0,0) circle [radius=0.16] node[white] {{\fontfamily{qag}\selectfont \tiny ID}};
        \draw[white, fill=white] (-0.0625,0.006) circle [radius=0.007];
    \end{tikzpicture}
    \hspace{-2.9mm}
}
\foreach \x in {A, B}{\expandafter\xdef\csname orcid\x\endcsname{\noexpand\href{https://orcid.org/\csname orcidauthor\x\endcsname}{\noexpand\orcidicon}}}

\newcommand{\iitm}{School of Mechanical and Materials Engineering, Indian Institute of Technology Mandi, Kamand 175075, India}

\newcommand{\gw}{G$_{0}$W$_{0}$}
\newcommand{\lz}{LiZnAs}
\newcommand{\sa}{ScAgC}
\newcommand{\rem}{Re$\epsilon_{\mathrm{M}}(\omega)$}
\newcommand{\iem}{Im$\epsilon_{\mathrm{M}}(\omega)$}
\newcommand{\remz}{Re$\epsilon_{\mathrm{M}}(0)$}

\newcommand{\eh}{\emph{e-h}}
\newcommand{\os}{oscillator strength}

\preprint{APS/123-QED}

\begin{document}


\title{Many-body \textit{ab initio} study of quasiparticles, optical excitations, and excitonic \\ properties in LiZnAs and ScAgC for photovoltaic applications}

\author{Vinod Kumar Solet \orcidA{}}
\email{vsolet5@gmail.com}
\author{Sudhir K. Pandey \orcidB{}}
\email{sudhir@iitmandi.ac.in}
\affiliation{\iitm}
\date{\today}

\begin{abstract}
In the search for novel photovoltaic (PV) solar materials, half-Heusler (HH) compounds have recently been regarded as especially promising due to their favorable electronic and optical properties. Using first-principles density-functional theory and many-body excited-state calculations, we study the quasiparticle band structure, as well as the optical and excitonic properties of two representative HH compounds, namely \lz\ and \sa, for PV applications. Our results reveal a direct band gap semiconducting behavior in \lz\ (\sa) with a value of $\sim$1.5 (1.0) eV under the accurate \gw\ method. The highest value of the imaginary part of the dielectric function is found to be $\sim$52 (87), 77 (87), 88 (91) using the independent-quasiparticle approximation, local field effects in the random-phase approximation, and electron-hole interaction within the Bethe-Salpeter equation, respectively. Both materials demonstrate a high refractive index, high absorption coefficients ($\sim$1.2-1.6 $\times$ 10$^6$ cm$^{-1}$), and low reflectivity (less than 40\%) within the active region of the solar energy spectrum. The triply degenerate bright excitons (exciton \textit{A}) at the main absorption peak, as well as a considerable number of bright excitonic states in the visible region, are observed; however, the excitons \os\ are comparatively weaker in \sa\ than in \lz. We further discuss the exciton character contributing to intense optical interband transitions and reveal that the direct optical band gap is associated to the loosely bound exciton \textit{A} state with binding energy of $\sim$45 (56) meV in \lz\ (\sa). Exciton \textit{A} is found to be highly localized (delocalized) in momentum (real) space, indicating the presence of Mott-Wannier-type excitons at band gap. Finally, we assess the solar efficiencies using the spectroscopic limited maximum efficiency (SLME) model and find SLME values of $\sim$32\% for \lz\ and $\sim$31\% for \sa\ at a $\sim$0.4 $\mu$m thin-film thickness. These findings highlight the significant role of excitons in solar energy absorption process and also suggest that both are highly suitable candidates for next-generation single-junction thin-film PV solar devices.
    
\end{abstract}

\maketitle

\section{Introduction} 
\setlength{\parindent}{3em}
The investigation of elementary excitations is one of the most effective research tools for understanding, predicting, and manipulating the properties of real materials for energy applications. For instance, excitons, which are many-body collective excitations associated with the mutual Coulomb \emph{attraction} between electron-hole (\eh) pairs, are commonly found in semiconductors and insulators \cite{bechstedt2016many}. The optical spectroscopy probes in these materials are largely governed by excitons \cite{fox2010optical}, making their inclusion crucial for accurately describing various technologies, including the generation of the photovoltaic (PV) effect for electricity production \cite{green2014emergence,schaller2004high}, solar-energy harvesting via photocatalytic water splitting into oxygen and hydrogen \cite{li2017review}, and the development of optoelectronic \cite{khan2008ultraviolet} and excitonic \cite{butov2017excitonic} devices. For example, the binding energy of photogenerated excitons is a critical quantity that influences key processes in solar cells, from light absorption to the direct generation of free charge-carriers \cite{grancini2013hot}. Their behavior are also crucial for efficient photogenerated free-charge transport and recombination \cite{savenije2014thermally}, which in turn influences the overall efficiency of the solar cell. On the other hand, PV device performance is also strongly influenced by the proper choice of materials; thus, researchers are making ongoing efforts to enhance the properties of existing materials or find novel alternatives to develop highly efficient devices.

To fully harness solar energy, selection criteria need to be identified to discover favorable PV materials. In particular, the properties that make any PV material highly efficient need to be well understood. At the computational level, novel solar materials can be aided by ensuring suitable band gaps and accurately calculating optical response functions. Many different theoretical methods exist for obtaining these properties, with density-functional theory (DFT)-based \emph{ab initio} methods being the most successful and widely used. However, a longstanding debate exists regarding the limitations of the single-particle DFT method, as it is exact for the ground state but provides only an approximate description of excited states, such as band gaps and optical properties \cite{hybertsen1985first,hybertsen1986electron,strinati1988application,onida2002electronic,aryasetiawan1998gw}. In recent years, most studies searching novel materials used in solar technology have assessed optical absorption properties and PV efficiency using one-particle DFT within the independent-particle (IP) approximation \cite{han2022ground,kangsabanik2020optoelectronic,nabi2024lead,
basera2020reducing,quan2022two,ghorui2024optoelectronic,dias2022first,sahni2020reliable,kangsabanik2019high,wang2021first,nabi2024lead,behera2024ternary,du2021cerium,sharan2022two}. However, optical absorption spectra obtained at this level show significant deviations from experiments \cite{bechstedt2016many}; therefore, IP-based properties can sometimes be misleading for accurately evaluating PV materials. Excitonic states at the band extrema (maxima and minima) are important for spectroscopic features in semiconductors; however, the IP framework completely neglects them \cite{bechstedt2016many}. This omission is a major source of inaccuracy in optical response calculation results compared to experiments \cite{bechstedt2016many,benedict1998optical,rohlfing1998electron}. Therefore, to achieve a more precise evaluation of PV materials, theoretical methods beyond single-particle DFT must be employed.

One of the most suitable approach to treat excited states in electronic materials is the Green’s function method \cite{hybertsen1986electron,onida2002electronic,strinati1988application}. Since excitons are collective excitations in the many-electron system, their \os\ cannot be accurately represented in the IP or one-particle DFT framework. This is because many-body effects, such as the quasiparticle (QP) and \eh\ interaction effects, are totally neglected in the independent \eh\ (free \eh\ pairs) approximation. To overcome these limitations, approaches that integrate DFT with many-body perturbation theory (MBPT) have become standard \cite{onida2002electronic} and are among the most versatile methods for studying charged and excitonic optical properties over the last few decades \cite{jiang2016gw,rohlfing1998electron,benedict1998optical,chang2000excitons}. This method utilizes QP ingredients and \emph{screened} Coulomb interactions W via one-body Green's function G (or propagator), leading to the so-called GW approximation, which describes individual quasielectron and quasihole excitations \cite{hybertsen1985first}. The excitons created in the light absorption process in a solar cell inherently involves two-particle properties; thus, the self-energy-corrected single-electron excitation picture, as described by the GW method at the purely electronic level, is not fully capable of capturing their behavior. Hence, an additional many-body treatment is necessary to analyze exciton behavior, which is achieved by including the \emph{screened} Coulomb interaction between holes in occupied states and electrons in empty states. These coupled \eh\ excitation states in a system are captured using two-body Green’s function approach based on Bethe-Salpeter equation (BSE) \cite{rohlfing1998electron}. Commonly, G is derived from the Kohn-Sham (KS) DFT and W is obtained through the dielectric function calculated from the random-phase approximation (RPA) \cite{hybertsen1986electron}. Such \textit{ab initio} calculations, properly accounting QP and excitonic effects, have been performed with a very high level of accuracy for optical absorption spectra in many materials, exhibiting excellent quantitative and qualitative agreement with experimental results \cite{solet2024,bechstedt2016many,rohlfing2000electron,albrecht1998ab,levine1989linear}. 

Filled tetrahedral semiconductors (FTSs) have long been recognized for their remarkable properties \cite{berger2020semiconductor,graf2011simple,solet2024}, making compound semiconductors an area of renewed technological interest \cite{dutta1997the,paskov1997refractive,dimroth2006high}. In particular, III-V-based zinc blende-like semiconductors such as GaAs have attracted significant attention for potential role in solar energy conversion \cite{green2021solar,yamaguchi2005multi,nelson2003the}. Within the FTSs family, ternary compounds (XYZ) such as half-Heusler (Nowotny–Juza) phases present opportunities to explore alternatives to III-V compound semiconductors \cite{wei1986electronic,roy2012half,wood1985electronic,belmiloud2016half,mavsek2007dilute,kalarasse2006optical}. Over time, half-Heusler (HH) semiconductors have exhibited intriguing characteristics, such as a vast variety of band gaps, piezoelectric coefficients and dielectric constants, and can thus serve as a promising platform for discovering novel materials for various sustainable energy applications \cite{graf2011simple,gruhn2010comparative,kieven2011preparation,belmiloud2016half,wood1985electronic,roy2012half,sahni2020reliable,solet2022first,shastri2020thermoelectric}. Furthermore, these materials are becoming increasingly relevant in modern research, owing to their diverse crystal structures, thermal stability and unique electronic, topological, magnetic, piezoelectric, and transport properties  \cite{graf2011simple,roy2012half,solet2023ab,shastri_FeVSb,pandey2021anab}. Structurally, these compounds comprise a combination of transition metals (X and Y) along with a p-block element (Z). They are easy to synthesize in various phases, such as metallic and insulating, and offer numerous possible combinations due to the vast selection of elements available for the three atomic sites \cite{legrain2018materials}. Notably, many HH semiconductors are predicted to have direct band gaps in the range of 1.0 to 2.0 eV \cite{wood1985electronic,gruhn2010comparative,belmiloud2016half} and thus making them highly relevant for PV solar applications.

In the search for the highly efficient solar cells, one could consider the I-II-V-type of \lz\ HH material. This HH system can be viewed as a zinc blende (ZnAS)$^{-}$ lattice with partially filled by He-like Li$^{+}$ interstitial sites \cite{wei1986electronic}. The (ZnAS)$^{-}$ sublattice is fundamentally similar to the well-known GaAs semiconductor, and indeed, \lz\ has been reported as a direct band gap semiconductor at both experimental \cite{kuriyama1987electrical,bacewicz1988preparation,kuriyama1994optical,kuriyama1996growth} and computational levels \cite{wei1986electronic,kacimi2014ii,gruhn2010comparative,chopra2019first,kalarasse2006optical,kamlesh2020effect,mehnane2012first,azouaoui2024comparative}. The experimental band gap of \lz\ ranges from 1.1 to 1.6 eV, which falls within the optimal band gap range required for solar cells. Its suitable band gap, along with its direct band gap nature, makes this compound highly promising for further evaluation in solar energy applications. Surprisingly, despite its favorable band gap, this compound has not been systematically screened for PV applications. For example, its optical properties have only been studied at the single-particle DFT level \cite{kalarasse2006optical,kamlesh2020effect,mehnane2012first,azouaoui2024comparative}, which is insufficient to fully realize its potential for solar cell applications. On the other hand, another HH material, \sa\ (I-III-IV type), which has a lattice parameter close to that of GaAs \cite{belmiloud2016half}, is also being considered for solar application properties investigations. The reason is that, in our previous study \cite{solet2022first}, \sa\ was identified as a suitable band gap material for solar cells, possessing a direct band gap nature with value of $\sim$1.03 eV using GW approximation calculation. Additionally, we studied its optical response properties at the IP approximation level. Therefore, both \lz\ and \sa\ still lack optical properties analysis using a more accurate many-body theoretical treatment, which is essential for assessing their full potential in solar technology. 

Therefore, the present work investigates the PV solar cell properties of \lz\ and \sa\ using DFT and many-body excited-state \emph{ab initio} calculations. To achieve this, we first calculate the electronic structure using PBE functional at the KS-DFT level, then apply the QP corrections using the \gw\ calculation for accurate band gap estimation, followed by optical and excitonic properties calculations. Both compounds exhibit direct band gap semiconducting behavior, with KS-DFT band gap values of $\sim$0.6 eV for \lz\ and $\sim$0.45 eV for \sa. After applying QP corrections, the band gaps increase to $\sim$1.5 and 1.0 eV, respectively. Furthermore, we compute the optical response properties at different levels of approximation: (i) independent quasiparticle (IQP) level without \eh\ interaction, (ii) local-field effects at the RPA level, and (iii) excitonic effects at the BSE level. The optical dielectric function obtained from all three methods exhibits the highest peak around the band gap edge in both materials. In this peak, a significant effect of local-field as well as excitonic corrections is observed for \lz, whereas both corrections are comparatively weaker in \sa. Alongside the largest \os\ of excitons below band gap energy, the intense peaks are also observed in the visible light region with considerable \os, suggesting that excitons play an important role in the light absorption process in solar cells. By analyzing the excitonic coefficients in reciprocal space, we find that the lowest-energy bright bound excitons related to main absorption peak in both materials are highly localized in this space. These bound excitons are triply degenerate (exciton \emph{A}) with binding energies of $\sim$45 meV in \lz\ and $\sim$56 meV in \sa. In real space, exciton \emph{A} is highly delocalized over several unit cells, indicating a Mott-Wannier character; hence positive sign of both materials for solar applications. Finally, we compute the spectroscopic limited maximum efficiency (SLME) and find the highest values of $\sim$32\% for \lz\ and $\sim$31\% for \sa\ at a $\sim$0.4 $\mu$m thickness. These values are significantly higher than that obtained for GaAs ($\sim$15\% at $\sim$0.5 $\mu$m) \cite{yin2015superior}. These results indicate that both \lz\ and \sa\ are highly promising solar materials for next-generation thin-film PV technology.

\section{THEORY AND METHODS} \label{sec:methods}

\subsection{\label{sec:2a}Optical spectroscopy including many-body effects}

One of the main purposes of this work is to obtain optical spectroscopic properties at an accurate theoretical levels by accounting for various approximations for both HH semiconductors used in PV applications. To solve the Bethe-Salpeter theory for optical dielectric function in our many-body perturbation calculations, the electronic structure of the compound is first needed, which is obtained via KS-DFT. It is well known that KS-DFT framework underestimates the band gap of semiconductors \cite{hybertsen1985first,jiang2016gw,bechstedt2016many}. Through the KS-DFT band structure, the QP energies (true single-particle excitation energies) and wave-functions are estimated using GW approximation calculations to account for electron-electron (\emph{e-e}) correlations \cite{hybertsen1986electron}. In this work, we have used a \gw\ (or one-shot GW) calculation. 
 
The BSE theory is described using two-particle (here electron and hole) propagators, which are four-point response functions (or density correlation function) that describe the simultaneous motion of both particles through the system, and can be written in a Dyson-like equation form \cite{onida2002electronic}:
\begin{eqnarray} \label{eq:L}
L = [1-L_{0}K]^{-1}L_{0}; \quad K = K^{\mathrm{x}} + K^{\mathrm{c}}                                       
\end{eqnarray}
where K$^{\mathrm{x}}$ and K$^{\mathrm{c}}$ are the exchange and direct Coulombic parts, respectively, of the BSE kernel K. The former is \emph{repulsive}, while the latter accounts for the \emph{attractive} nature of the \eh\ interaction. L is the \eh\ correlation function, and L$_{0}$ is its interaction-free version; \textit{i.e.}, when (quasi)electron and (quasi)hole do not interact (K = 0), L reduces to L$_{0}$. In direct (\textbf{r}) space, Eq.~\hyperref[eq:L]{\eqref{eq:L}} represents the four-point function, whereas in the basis of \eh\ transitions or the products of hole states ($v\textbf{k}$) and electron states ($c\textbf{k}$) [$vc\textbf{k}$], it can be treated as a matrix form with L = L$_{vc\textbf{k}, v^{\prime}c^{\prime}\textbf{k}^{\prime}}$. The full (reducible) many-body two-point polarization function
\begin{eqnarray} \label{eq:P}
\textbf{P}(\textbf{r},\textbf{r}^{\prime};\omega) = -i L(\textbf{r},\textbf{r},\textbf{r}^{\prime},\textbf{r}^{\prime};\omega)                                       
\end{eqnarray}
is a diagonal element of the response function L and is used to obtain the macroscopic dielectric function ($\epsilon_{\mathrm{M}}$) in terms of the inverse of the microscopic dielectric function ($\epsilon$) \cite{onida2002electronic},
\begin{eqnarray} \label{eq:epsilon}
\epsilon (\textbf{r},\textbf{r}^{\prime};\omega) = \delta(\textbf{r},\textbf{r}^{\prime}) - \int v_{c}(\textbf{r},\textbf{r}_{1}) \textbf{P}(\textbf{r}_{1},\textbf{r}^{\prime};\omega)d\textbf{r}_{1}.                                       
\end{eqnarray}
Where $v_{c}(\textbf{r},\textbf{r}^{\prime})$ = 1/$|\textbf{r}-\textbf{r}^{\prime}|$ is the bare Coulomb potential. Further, an eigenvalue problem for an effective \eh\ interacting (excitonic) Hamiltonian (H$^{\mathrm{exc}}$),
\begin{eqnarray}\label{eq:exc}
H^{\mathrm{exc}} A^{\lambda} = E^{\lambda} A^{\lambda} 
\end{eqnarray}  
must be solved for L by obtaining the eigenvalues E$^{\lambda}$ of the \eh\ interacting pair $\lambda$ ($\lambda$th exciton) with corresponding eigenvectors A$^{\lambda}$
\begin{eqnarray} \label{eq:lexc}
L_{vc\textbf{k}, v^{\prime}c^{\prime}\textbf{k}^{\prime}} = i \sum_{\lambda} \frac{A^{\lambda}_{vc\textbf{k}}[A^{\lambda}_{v^{\prime}c^{\prime}\textbf{k}^{\prime}}]^{*}}{\omega - E^{\lambda}}.
\end{eqnarray}  

Furthermore, the H$^{\mathrm{exc}}$ is required to obtain \textbf{P} in Eq.~\hyperref[eq:epsilon]{\eqref{eq:epsilon}}, which, for spin-unpolarized case (spin-singlet excitons), is written as
\begin{eqnarray} \label{eq:ham}
H^{\mathrm{exc}} = H^{\mathrm{diag}} + 2H^{\mathrm{x}} + H^{\mathrm{c}}.
\end{eqnarray}
Besides the diagonal part H$^{\mathrm{diag}}$, representing excitation of independent (quasi)particles [I(Q)P], the \emph{repulsive} exchange term, H$^{\mathrm{x}}$ mainly originates from the \emph{unscreened} (bare) Coulomb interaction $v_{c}$ between hole and electron, and the \emph{attractive} term of \eh\ correlation, H$^{\mathrm{c}}$ is mediated by the statically ($\omega$ = 0) \emph{screened} version of $v_{c}$ (denoted by W) due to the dielectric medium. Their expressions are written as \cite{gulans2014exciting},
\begin{equation} \label{eq:bseH}
\begin{aligned}
    &H^{\mathrm{diag}}_{vc\mathbf{k}, v^{\prime}c^{\prime}\mathbf{k}^{\prime}} = E_{vc\textbf{k}}\delta_{c,c^{\prime}} \delta_{v,v^{\prime}} \delta_{\mathbf{k},\mathbf{k}^{\prime}}, \\
    &H^{\mathrm{x}}_{vc\mathbf{k}, v^{\prime}c^{\prime}\mathbf{k}^{\prime}} = \int d^{3}\mathbf{r} \, d^{3}\mathbf{r}^{\prime} \, \phi_{v\mathbf{k}}(\mathbf{r}) \phi_{c\mathbf{k}}^{*}(\mathbf{r}) \, \bar v_{c}(\mathbf{r}, \mathbf{r}^{\prime}) \, \\
    &\quad \times \phi_{v^{\prime}\mathbf{k}^{\prime}}(\mathbf{r}^{\prime}) \phi_{c^{\prime}\mathbf{k}^{\prime}}^{*}(\mathbf{r}^{\prime}), \\
    &H^{\mathrm{c}}_{vc\mathbf{k}, v^{\prime}c^{\prime}\mathbf{k}^{\prime}} = -\int d^{3}\mathbf{r} \, d^{3}\mathbf{r}^{\prime} \, \phi_{v\mathbf{k}}(\mathbf{r}) \phi_{c\mathbf{k}}^{*}(\mathbf{r^{\prime}}) \, \\
    &\quad \times W(\mathbf{r}, \mathbf{r}^{\prime};\omega = 0) \, \phi_{v^{\prime}\mathbf{k}^{\prime}}^{*}(\mathbf{r}) \phi_{c^{\prime}\mathbf{k}^{\prime}}(\mathbf{r}^{\prime}).
\end{aligned}
\end{equation}
Here E$_{vc\textbf{k}}$ = $\varepsilon_{c\mathbf{k}}^{qp} - \varepsilon_{v\mathbf{k}}^{qp}$ are the self-energy-corrected QP energies difference of empty electron states and filled hole states. But, to obtain these energies, we just apply here scissor correction on the conduction KS energies. To describe transitions between independent quasielectron and quasihole states within the IQP, only the H$^{\mathrm{diag}}$ term in Eq.~\hyperref[eq:ham]{\eqref{eq:ham}} is switched on, and the optical spectra are then obtained by solving Eq.~\hyperref[eq:P]{\eqref{eq:P}} without including local field effects (LFEs) \cite{onida2002electronic,hybertsen1986electron}. By additionally incorporating the \eh\ exchange interaction H$^{\mathrm{x}}$ term, the spectra are retrieved within the RPA, now effectively accounting for the effects of microscopic fields; \textit{i.e.}, the LFEs that locally counteract the external field effect. In this context, the distinction between IQP and RPA arises solely from the inclusion of LFEs. Note that the Coulomb potential $\bar v_{c}$ used in exchange Hamiltonian is defined without its long-range component at \textbf{G} = 0 \cite{onida2002electronic}. Finally, by also switching the Coulombic \eh\ \emph{attraction} H$^{\mathrm{c}}$ term, the full BSE Hamiltonian in Eq.~\hyperref[eq:ham]{\eqref{eq:ham}} captures the excitonic effect in the system. 

In a dielectrically polarized medium, \emph{screened} potential W is obtained using Eq.~\hyperref[eq:epsilon]{\eqref{eq:epsilon}} through RPA,
\begin{eqnarray} \label{eq:W}
W(\mathbf{r}, \mathbf{r}^{\prime};\omega = 0) = \int \epsilon^{-1}(\mathbf{r}, \mathbf{r}_{1};\omega = 0) v_{c}(\mathbf{r_{1}}, \mathbf{r}^{\prime}) d\mathbf{r}_{1}. \quad\quad
\end{eqnarray}

The optical absorption or more precisely, the imaginary part of $\epsilon_{\mathrm{M}}$ [\iem] is obtained by only considering the resonant part of BSE Hamiltonian matrix within the Tamm-Dancoff approximation (TDA) \cite{vorwerk2019bethe}. To probe optical absorption spectroscopy, the wave vector of photon is very small compared to crystal lattice dimension. Thus, in the long wavelength limit \textbf{q}$\to$ 0, using the Fourier transform of Eq.~\hyperref[eq:epsilon]{\eqref{eq:epsilon}}, one obtains \cite{onida2002electronic}:
\begin{eqnarray} \label{eq:iem}
\mathrm{Im}\epsilon_{\mathrm{M}}(\omega) = \lim_{\textbf{q}\to 0}\frac{8\pi^{2}}{Vq^{2}}\sum_{\lambda}\Bigg|\sum_{vc\textbf{k}} A^{\lambda}_{vc\textbf{k}} \tilde{\rho}_{vc\textbf{k}}(\textbf{q})\Bigg|^{2}\delta(\omega-E^{\lambda}) \quad \quad
\end{eqnarray}
where V is the crystal volume, and $\tilde{\rho}_{vc\textbf{k}}$(\textbf{q}) = $\langle\phi_{v\textbf{k}-\textbf{q}}(\textbf{r})|e^{-i\textbf{q.r}}|\phi_{c\textbf{k}}(\textbf{r})\rangle$ is related to the \os\ of IQP transitions. If these independent transitions are combined through excitonic states A$^{\lambda}$, that is, the square modulus of Eq.~\hyperref[eq:iem]{\eqref{eq:iem}}, then they are typically referred to as the dipole \os s of exciton $\lambda$. In our IQP, RPA and BSE calculations, $\phi_{i}$ are the KS states. 

In the IQP approximation, the Eq.~\hyperref[eq:iem]{\eqref{eq:iem}} reduces to the well-known interband formula,
\begin{eqnarray} \label{eq:iem-qp}
\mathrm{Im}\epsilon_{\mathrm{M}}(\omega) = \lim_{\textbf{q}\to 0}\frac{8\pi^{2}}{Vq^{2}}\Bigg|\sum_{vc\textbf{k}} \tilde{\rho}_{vc\textbf{k}}(\textbf{q})\Bigg|^{2}\delta(\omega-E_{vc\textbf{k}}) \quad \quad
\end{eqnarray}

Once the BSE is solved, the character and distributions of the \eh\ wave function in reciprocal space can also be analyzed through the excitonic weights of a transition at each \textbf{k} point in the valence and conduction bands as 
\begin{subequations} \label{eq:wave-reci}
\begin{eqnarray}   
w_{v\textbf{k}}^{\lambda} &=& \sum_{c}|A_{vc\textbf{k}}^{\lambda}|^{2},  \label{eq:wave-reci-a} \\
w_{c\textbf{k}}^{\lambda} &=& \sum_{v}|A_{vc\textbf{k}}^{\lambda}|^{2}.  \label{eq:wave-reci-b}
\end{eqnarray}
\end{subequations}

Similarly one can also get insight into the localization (or wave functions) of exciton in real space by a linear combination of the Bloch functions for occupied and unoccupied bands weighted by exciton eigenvector \cite{bechstedt2016many}
\begin{eqnarray} \label{eq:wave-real}
\Psi_{\lambda}(\textbf{r}_{e}, \textbf{r}_{h}) = \sum_{vc\textbf{k}} A^{\lambda}_{vc\textbf{k}} \phi_{c\textbf{k}}(\textbf{r}_{e}) \phi_{v\textbf{k}}(\textbf{r}_{h}).
\end{eqnarray}

\subsection{\label{sec:2d}Computational details}

Our many-body \emph{ab initio} calculations are carried out using the full-potential all-electron \textit{exciting} code \cite{vorwerk2019bethe}, which employs the (linearized) augmented plane-wave + local orbital basis set in the KS equation to compute both occupied and unoccupied states. The KS ground-state part of MBPT is obtained using the Perdew-Burke-Ernzerhof (PBE) functional \cite{perdew1996generalized} with a \textbf{k}-point grid of $12\times 12\times 12$ and a plane-wave cutoff of R$_{\mathrm{MT}}$$|\textbf{G}+\textbf{K}|_{\mathrm{max}}$ = 7.0. After constructing the one body Green's functions and \emph{screened} Coulomb interaction using KS orbitals and energies obtained from PBE as input for the single-shot \gw\ calculations, the QP correction is then computed at 12 frequency-grid using different \textbf{k}- and \textbf{q}-point meshes as well as different number of empty states, as shown in table~\ref{tab:gap}.  

We use KS wave functions and scissor-corrected KS eigenvalues as a starting point for obtaining the optical spectroscopy using IQP and RPA, as well as including \eh\ \emph{attraction} within BSE under TDA. The scissor correction values of 0.9 eV and 0.55 eV for \lz\ and \sa, respectively, are applied to further open the gap towards the \gw\ gap value. For all optical response calculations, a $5\times 5\times 5$ \textbf{k}-point grid and 50 empty states are employed. For screening calculations, a $5\times 5\times 5$ \textbf{q}-grid and 100 empty states are used. To include LFEs in the exchange part of the RPA calculations, an energy cutoff of 3.0 Hartree is set. To construct the BSE Hamiltonian in the basis of conduction and valence band states (transition space), the highest 12 occupied and first 20 unoccupied bands are considered. Finally, all optical spectra are plotted using a Lorentzian broadening value of 50 meV.

\begin{figure}[ht]
\includegraphics[width=7.9cm, height=4.6cm]{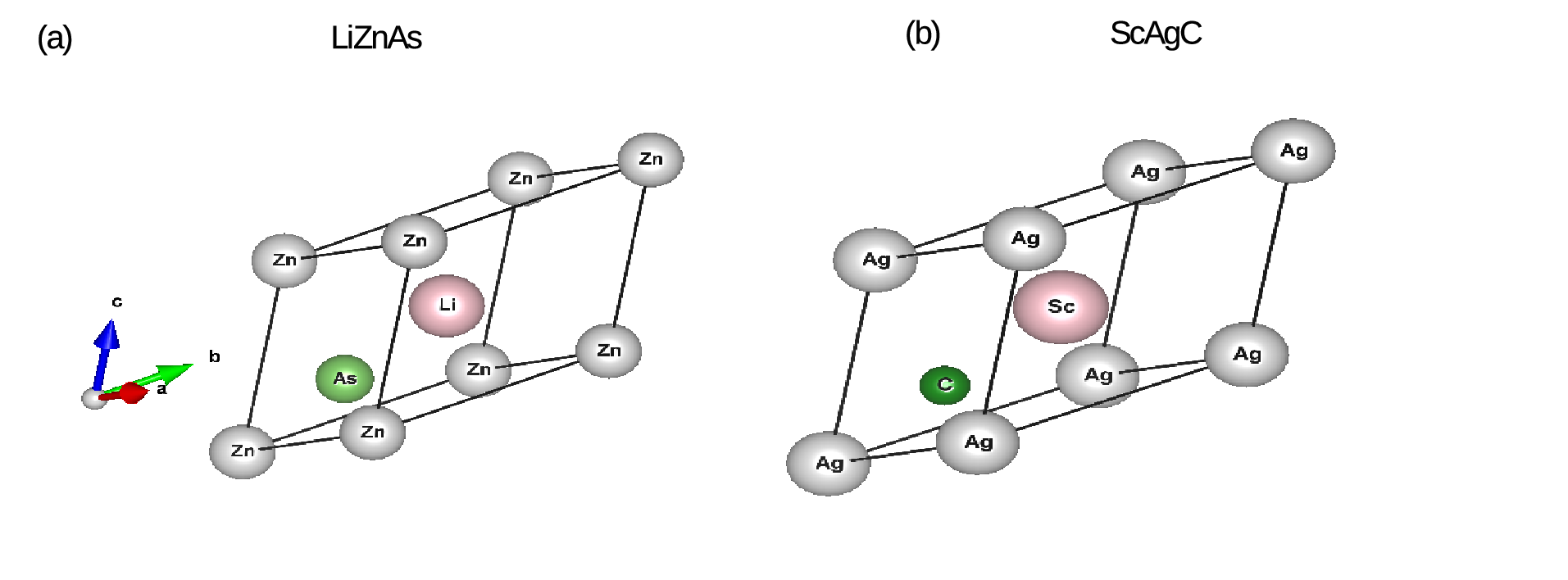} 
\caption{Crystal structure for (a) \lz\ and (b) \sa\ half-Heusler compounds.} 
\label{fig:cry}
\end{figure}

Both HH compounds crystallize in the space group F$\bar{4}$3m (number 216) of the face-centered cubic system. The atomic arrangements of the structures in the primitive cell for both HH materials, obtained using the VESTA software \cite{momma2011vesta}, are shown in Fig.~\hyperref[fig:cry]{\ref{fig:cry}}. The primitive unit cell is used in all calculations, and the Wyckoff positions of the atoms are as follows: Zn (Ag) at 4a (0.0, 0.0, 0.0), Li (Sc) at 4b (0.5, 0.5, 0.5), and As (C) at 4c (0.25, 0.25, 0.25) in the \lz\ (\sa) compound \cite{kuriyama1994optical,solet2022first}. The lattice parameter values used are 5.94 \AA\ for \lz\ \cite{kuriyama1994optical} and 5.59 \AA\ for \sa\ \cite{solet2022first}. The default muffin-tin radius (R$_{MT}$) values have been used for all atoms.

\section{Results and discussion} \label{sec:result}
\subsection{\label{sec:3a}Electronic structure and quasiparticle effect}

\begin{figure*}[ht]
\includegraphics[width=13.7cm, height=6.7cm]{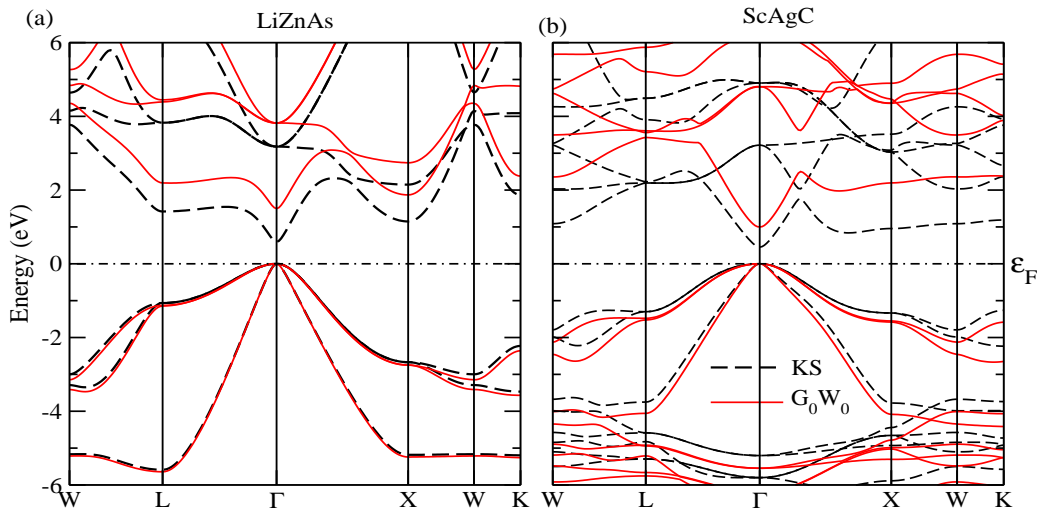} 
\caption{The Kohn-sham (KS) and \gw\ quasiparticle band structures using PBE functional for (a) \lz\ and (b) \sa\ half-Heusler compounds. Fermi energy $\varepsilon_{\mathrm{F}}$ (at zero energy) is at top of the valence band.} 
\label{fig:disp}
\end{figure*}

We start our analysis by obtaining the electronic band structures based on different methods. Figure~\hyperref[fig:disp]{\ref{fig:disp}} compares the KS-PBE and QP band structures along high-symmetric path for both \lz\ and \sa\ HH compounds. In both materials, a direct band gap is observed at the $\Gamma$-point, with KS-PBE calculations yielding values of $\sim$0.6 eV for \lz\ and $\sim$0.45 eV for \sa. As commonly noted in table~\ref{tab:gap}, KS-PBE underestimates experimental band gaps, which have been reported for \lz\ compound at 1.1 eV from electrical resistivity measurement \cite{kuriyama1987electrical}, 1.25 eV from optical absorption edge and photoconductivity measurement \cite{bacewicz1988preparation}, and 1.51 and 1.61 eV from photoluminescence and optical absorption measurements (for single crystal) using a scanning spectrophotometer \cite{kuriyama1994optical}. Unfortunately, no experimental studies have been conducted on \sa\ so far. By considering many-body effects, the QP-corrected band gap at the \gw\ level is found to be $\sim$1.5 eV for \lz\ and $\sim$1 eV for \sa, indicating a self-energy correction of more than twice the KS gap in both compounds. This significant QP correction arises purely from enhanced \emph{e-e} interactions, emphasizing the strong influence of Coulombic screening effects. A convergence test for the \gw\ band gap with respect to \textbf{k}- and \textbf{q}-point sampling, as well as the number of empty states, is also performed. The results are shown in table~\ref{tab:gap}, where no considerable change in the \gw\ gap value is observed upon varying the wavevectors and unoccupied states.

By aligning the KS and QP band structures at the valence band (VB) maxima (VBM), we gain a clearer perspective on how self-energy corrections affect the valence and conduction regions. In \lz, the occupied states in the region of L-$\Gamma$-X are identically reproduced by the KS-DFT and \gw\ calculations. In contrast, for \sa, we observe a slight lowering of the occupied band energies in this \textbf{k}-point region (except around $\Gamma$) when the QP effect is considered. This suggests that self-energy corrections to band dispersions are comparatively larger in \sa\ than in \lz. Due to this correction, the fundamental nature of the band gap remain consistent between PBE and \gw\ approaches. The band structure of \lz\ has shown great similarity with GaAs semiconductor \cite{wei1986electronic}. The obtained \gw\ gap value of \lz\ is almost same as the value observed for single-crystal \lz\ in optical absorption measurements \cite{kuriyama1994optical}. Moreover, the obtained \gw\ band gaps of both HH alloys are in the desired range for solar energy absorption, and thus they deserve for further analysis of the excitonic features in optical parameters for efficient PV performance.

\begin{figure*}[ht]
\includegraphics[width=17.2cm, height=12.7cm]{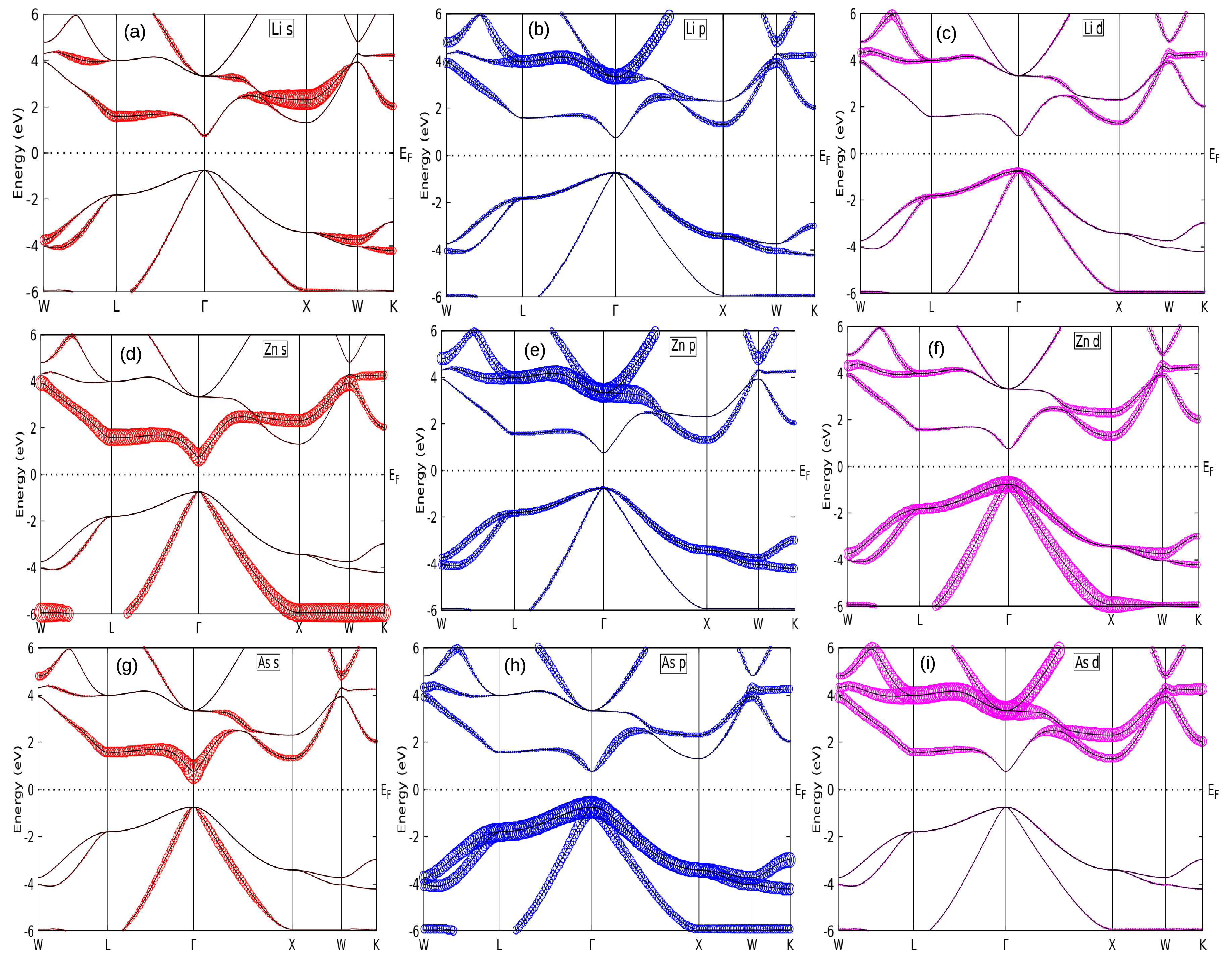} 
\caption{The orbital-resolved contribution of each species in \lz\ projected with color on the QP band structure. The orbital characters shown are: (a) Li \textit{s}; (b) Li \textit{p}; (c) Li \textit{d}; (d) Zn \textit{s}; (e) Zn \textit{p}; (f) Zn \textit{d}; (g) As \textit{s}; (h) As \textit{p}; (i) As \textit{d}.} 
\label{fig:bandch-li}
\end{figure*}

\begin{figure*}[ht]
\includegraphics[width=17.2cm, height=11.7cm]{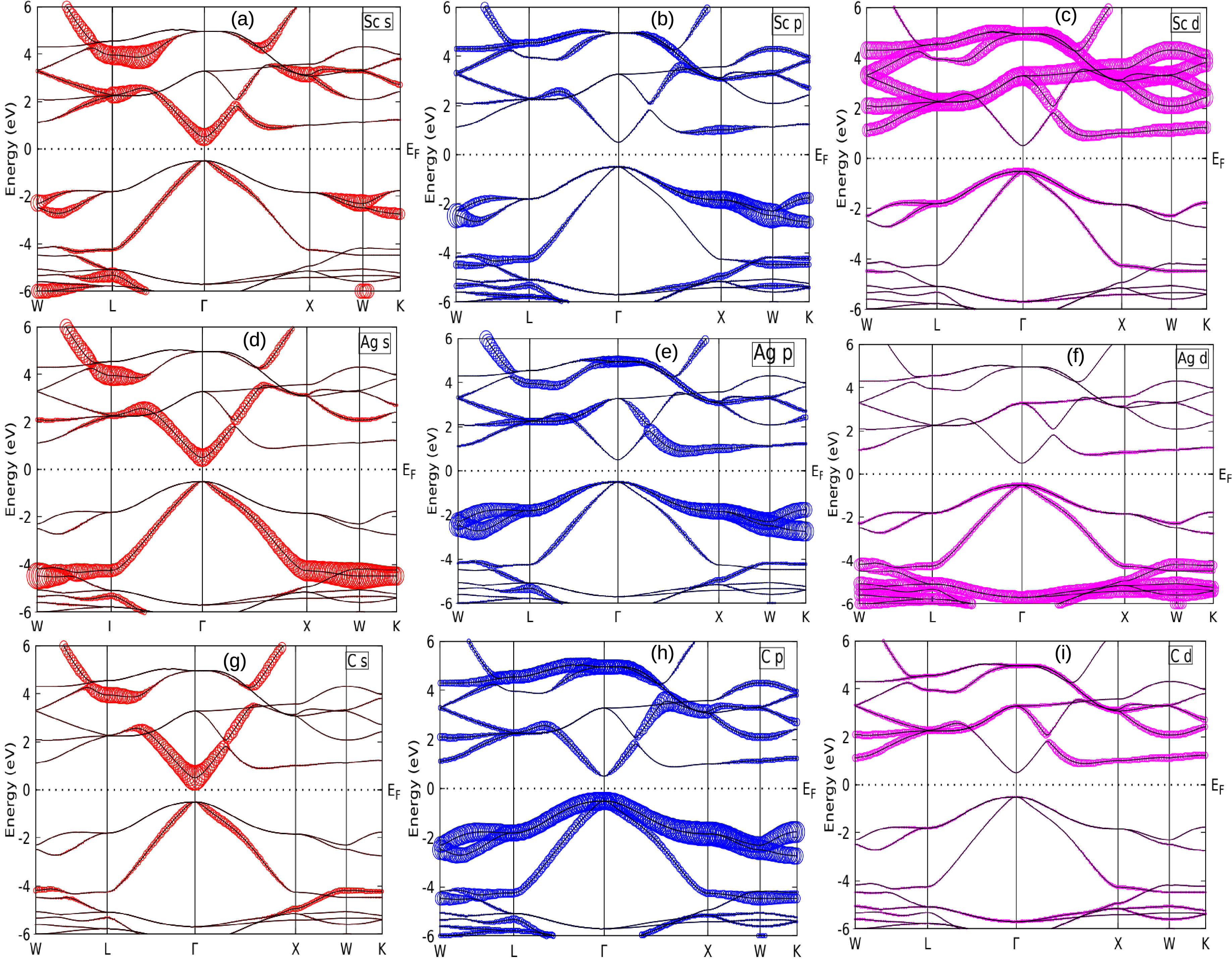} 
\caption{The orbital-resolved contribution of each species in \sa\ projected with color on the QP band structure. The orbital characters shown are: (a) Sc \textit{s}; (b) Sc \textit{p}; (c) Sc \textit{d}; (d) Ag \textit{s}; (e) Ag \textit{p}; (f) Ag \textit{d}; (g) C \textit{s}; (h) C \textit{p}; (i) C \textit{d}.} 
\label{fig:bandch-sc}
\end{figure*}

In the next step of our discussion, the orbital-resolved character of different atoms projected onto the band structure in both compounds (see Fig.~\hyperref[fig:bandch-li]{\ref{fig:bandch-li}} for \lz\ and Fig.~\hyperref[fig:bandch-sc]{\ref{fig:bandch-sc}} for \sa) is investigated. This analysis is important for several reasons. First, understanding the nature of different electronic states (or orbitals) in the valence and conduction band regions helps in interpreting and predicting optical properties by determining the selection rules for optical transitions. Additionally, the orbital character of bands provides valuable insights into exciton transport behavior in the respective bands. Since tuning excitonic properties is crucial for solar cell materials, this understanding plays a key role in optimizing their performance.

In both materials, it is apparent that the VBM exhibits a strong $p$-like atomic character, primarily due to the p-block elements. An additional contribution to the VBM receives from the Zn $d$ state, with a minor extent from the Li $d$-orbital in \lz\ and minimal contributions from the $d$-states of Sc and Ag in \sa. This suggests that the $p$-states of nonmetal p-block elements in both materials control largely the transport around the VBM in solar cells. The conduction band (CB) minima (CBM) is derived mostly from the $s$-like state of the As atom, with a considerable contribution from the Zn $s$-like orbital as well. Similarly, in \sa, the CBM has a strong $s$-like character, with relatively smaller contributions from the $s$-states of Sc and Ag. We acknowledge the selection rules for dipole transitions, which allow $p\rightarrow s$ transition \cite{fox2010optical}. Therefore, one can conclude that most transitions from occupied to unoccupied states in both materials are electric-dipole allowed, and as a result, strong optical absorption is expected. This highlights the crucial role of the $s$ and $p$ states in both materials in determining the \eh\ formation process associated with optical transitions. Besides these band extrema, orbitals with significant contribution above the CBM and below the VBM along the studied high-symmetry path are also critically responsible for the solar energy conversion process. The character of these orbitals further provides promising insights for identifying the fingerprints of excitons that contribute to the optical spectra, which will be thoroughly discussed in the next section.  

\begin{table}[t]
\caption{\label{tab:gap}The lattice constants and fundamental band gaps (in eV) from PBE and \gw\ calculations for both materials. The convergence of the \gw\ gap is tested with respect to \textbf{k} and \textbf{q} points (both meshes are the same) and the number of empty states.}
\begin{ruledtabular}
\begin{tabular}{lccccc}
 & LiZnAs & ScAgC \\
\hline
Lattice constant (\AA) & 5.94 & 5.59 \\
PBE gap & 0.6 & 0.45 \\
\hline
\multirow{1}{*}{\gw\ gap at 100 empty states}\\ 
$4 \times 4 \times 4$ mesh & 1.522 & 0.998 \\
$5 \times 5 \times 5$ mesh & 1.504 & 1.012 \\
$6 \times 6 \times 6$ mesh & 1.498 & 1.030 \\
\hline
\multirow{1}{*}{\gw\ gap at $6 \times 6 \times 6$ mesh}\\ 
 200 empty states & 1.501 & 1.067 \\
 300 empty states & 1.496 & 1.052 \\
\hline
Experimental gap & 1.1--1.61 & Not available \\
\end{tabular}
\end{ruledtabular}
\end{table}

\begin{figure*}[ht]
\includegraphics[width=18.0cm, height=9.5cm]{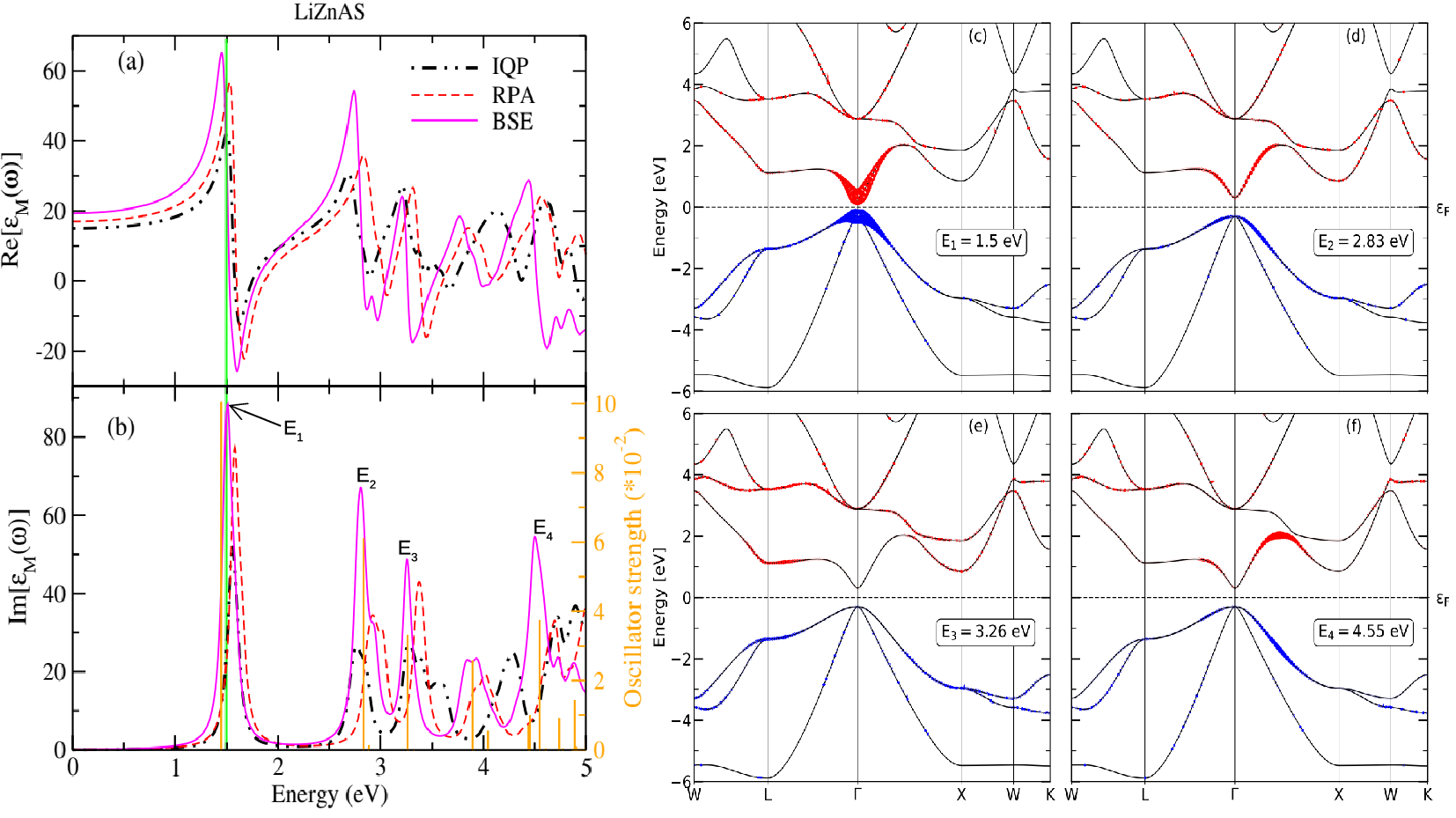} 
\caption{(a) Real and (b) imaginary parts of the dielectric function for \lz\ calculated using methods based on the IQP, RPA, and BSE, where the optical (direct) \gw\ band gap is marked by a vertical solid green line. The vertical orange lines in (b) represent the \os\ and position of the excitation, indicate the solution of individual BSE exciton. (c-f) Electron-hole coupling pair (exciton) coefficients in reciprocal space are represented by blue and red circles, superimposed on the DFT band structure for intense peaks at different transition energies, as marked in (b). The radius of each circle is proportional to the square of the amplitude of the $\lambda$th excitonic wave function ($|A^{\lambda}_{vc\textbf{k}}|^{2}$). } 
\label{fig:ex-li}
\end{figure*}

\begin{figure*}[ht]
\includegraphics[width=18.0cm, height=9.5cm]{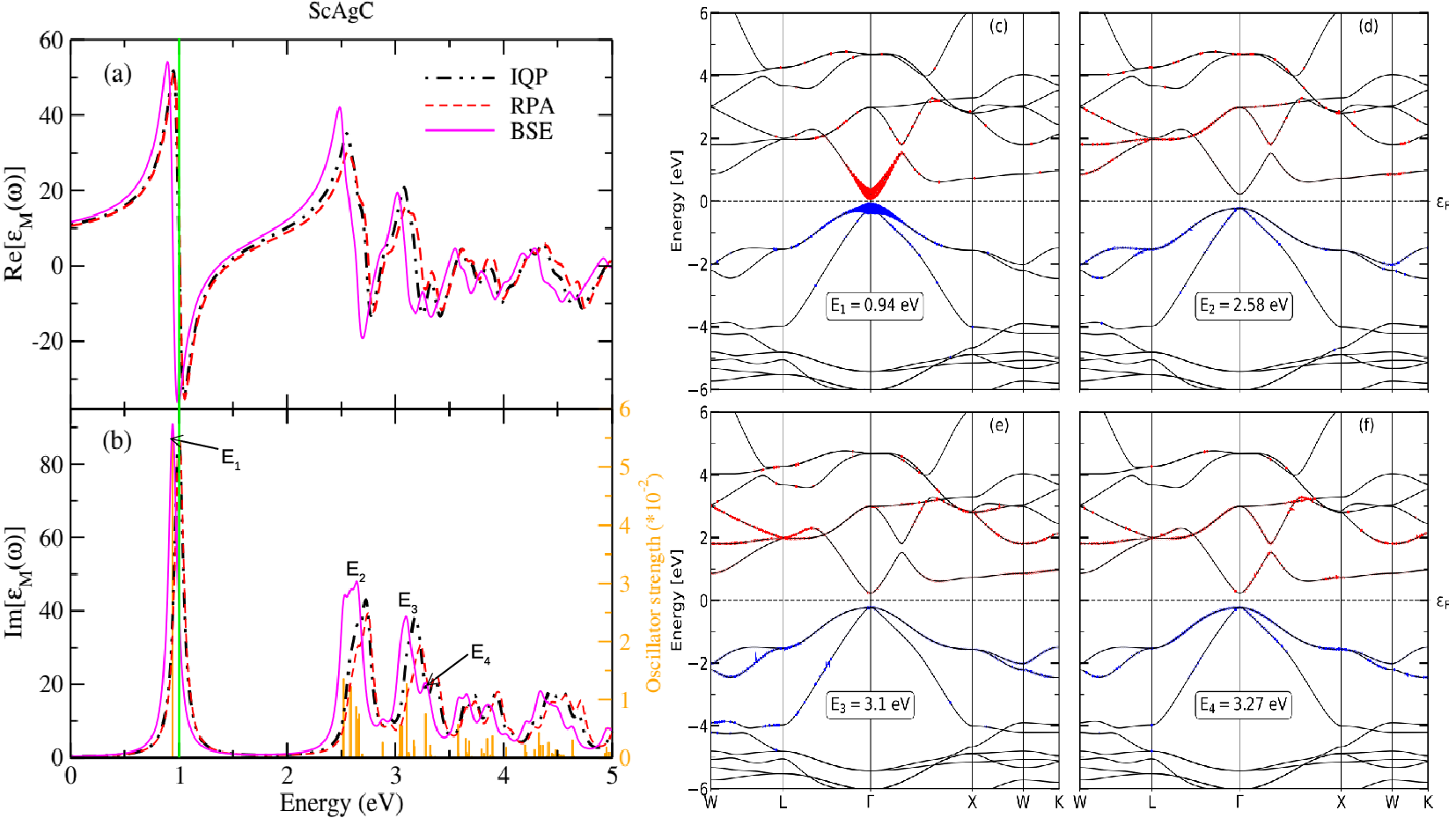} 
\caption{Plots analogous to those in Fig.~\ref{fig:ex-li} but for the \sa\ hH material. (c)-(f) The excitonic coefficients in reciprocal space related to intense intense peaks at different transition energies, as marked in (b): (c) E$_1$ = 0.94 eV; (d) E$_2$ = 2.58 eV; (e) E$_3$ = 3.1 eV; (f) E$_4$ = 3.27 eV.  }
\label{fig:ex-sc}
\end{figure*}

\begin{figure*}[ht]
\includegraphics[width=13.7cm, height=7.4cm]{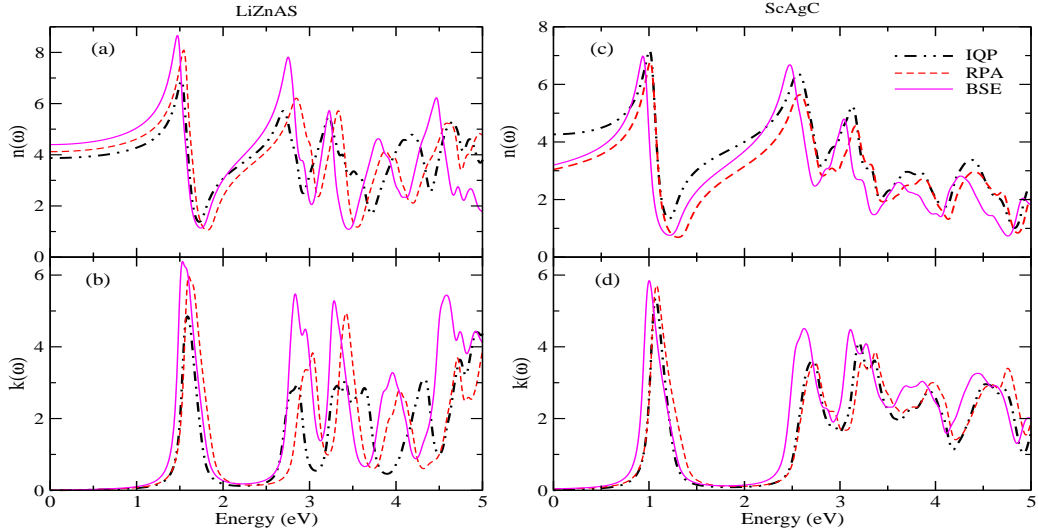} 
\caption{The calculated real n($\omega$) and imaginary k($\omega$) parts of refractive index using methods based on IQP, RPA and BSE for \lz\ in (a, b) and for \sa\ in (c, d).  } 
\label{fig:refra}
\end{figure*}

\subsection{\label{sec:3b}Optical response and exciton analysis in reciprocal space}

Following the electronic structure discussed in the preceding section, we determine the linear optical properties resulting from the interaction of photons with electrons under electromagnetic radiation and analyze the behavior of excitons in spectra by including many-body effects. We examine the effects that go beyond the exchange-correlation functional, primarily including QP correction within IQP approximation, the role of the LFEs within the RPA and excitonic corrections within the BSE. This analysis is completely based on interband transitions to produce spectral features in reciprocal space. By using of Kramers-Kronig transformation of \iem\ in Eq.~\hyperref[eq:iem]{\eqref{eq:iem}}, the real part of $\epsilon_{\mathrm{M}}$ [\rem] and other optical parameters such as refractive index, absorption coefficients, and reflectivity can be easily computed \cite{solet2022first}.

In the first part of PV properties discussion, we compare the macroscopic dielectric function $\epsilon_{\mathrm{M}}$ obtained via IQP, RPA and BSE methods and discuss the exciton behavior related to large absorption peaks for both HH compounds. The results are shown in Fig.~\ref{fig:ex-li} for \lz\ and in Fig.~\ref{fig:ex-sc} for \sa. Photon energy dependent \rem\ and \iem\ are plotted in panels (a) and (b), respectively, of both figures for these compounds. The vertical solid green line in these figures represents the \gw\ band gap of the respective materials. In both materials, various sharp, intense peaks are observed in the studied energy range, indicating strong photo-induced transitions between states of holes and electrons. The main peak of both real and imaginary parts in both materials is located around band gap edge, which makes it interesting feature in the context of solar cells.

The value of ion clamped static macroscopic dielectric constant $\epsilon_{\infty}$ = \remz\ for \lz\ increases slightly when we include LFEs and then \eh\ \emph{attraction} into IQP picture, whereas the opposite behavior is observed for \sa. The values are $\sim$15 (18.2), 17 (9.3), and 19.4 (10.2) for \lz\ (\sa) using IQP, RPA, and BSE, respectively. These large values indicate that the Coulombic effects due to dielectric screening between holes and electrons are small, suggesting that the exciton binding energies in these materials are expected to be low, on the order of a few meV. The decreasing trend of $\epsilon_{\infty}$ with local field correction has been generally observed in many semiconductors, such as Si, GaAs, SiC, and AlP, as well as in insulators like diamond \cite{gajdovs2006linear}. In addition to this, large effect of LFEs on the RPA spectra is observed for \lz\ compared to \sa. Also, the value of first main peak of \rem\ increases for \lz, while it decreases for \sa. When incorporating the \eh\ interaction picture using BSE formalism, the magnitude of the peaks increases significantly for \lz\ compared to \sa. 

In Fig.~\hyperref[fig:ex-li]{\ref{fig:ex-li}(b)} for \lz, the main peak of \iem\ is found to be near the band gap edge, with intensity values of $\sim$52, 77, and 88 obtained from IQP, RPA and BSE methods, respectively. This indicates that the peak value significantly increases due to local-field correction. Meanwhile, a strong excitonic feature is also present in this highest absorption peak. In Fig.~\hyperref[fig:ex-sc]{\ref{fig:ex-sc}(b)}, the corresponding values obtained from these methods are $\sim$87, 87, and 91 for \sa. This indicates that LFEs have no effect on the creation of the main absorption peak in \sa, while only a slight change in magnitude is observed when \eh\ \emph{attraction} is included. It means that this peak is no longer be affected by the presence of excitonic states. It seems from these results that the \eh\ interaction is stronger in \lz\ compared to \sa. One thing to note in both materials is that the energy position of this peak remains almost unchanged when LFEs are included over IQP, whereas it shifts towards a lower energy side due to the excitonic effects. Exciton position related to E$_{1}$ peak lies just below the band gap energy, serving as strong evidence of the presence of bound excitons in these materials. 

In additions to the excitation at band gap, further peaks are located at higher energies in both alloys. For example, the E$_{2}$ peak, located in the visible energy region in both materials, experiences a more significant redistribution of the \os\ compared to other peaks marked in the figures. This is attributed to the strong excitonic effects, indicating  that excitons generated by visible light photons can play a crucial role in charge carrier generation in the neutral \emph{p} and \emph{n} regions of \emph{pn} junction solar cells. In the energy region of $\sim$1.8-2.5 eV for \lz\ and $\sim$1.3-2.2 eV for \sa, the observed values drop to near zero, implying that photons within these energies do not effectively contribute to generating the PV effect and go to waste. At higher energies, a more considerable redistribution of the \os\ is observed in the creation of the E$_{4}$ peak compared to E$_{3}$ in \lz. In \sa, despite the peaked shape of E$_{3}$ and E$_{4}$, they do not exhibit large excitonic behavior, as their strength remains largely unchanged compared to the IQP spectrum. A more clear understanding of the BSE spectra can be obtained by analyzing the origin of excitons in the various studied peaks for both PV materials.

The excitonic features in spectrum can be analyzed through the mixing in the IQP transition at various \textbf{k} points in the Brillouin zone, weighted by the coefficient of a particular exciton $\lambda$, A$^{\lambda}_{vc\textbf{k}}$, via Eq.~\hyperref[eq:wave-reci]{\eqref{eq:wave-reci}}. Therefore, the character of an exciton is directly related to the analysis of excitonic coefficients. 

In this manner, we first analyze the intensity of the each exciton \textit{i.e.}, \os\ located at energy E$^{\lambda}$, which is plotted using vertical orange bars in Fig.~\hyperref[fig:ex-li]{\ref{fig:ex-li}(b)} for \lz\ and Fig.~\hyperref[fig:ex-sc]{\ref{fig:ex-sc}(b)} for \sa. Excitons with negligibly small magnitude are known as dark; otherwise, they are said to be bright. The \os\ of excitons is comparatively larger in \lz\ than in \sa. In both materials, only optically bright excitons, three in number, are found in the band gap energy region, which is a direct signature of the formation of the largest peak in these materials. Interestingly, all three excitons are degenerate in both materials, with an excitation energy of 1.452 eV in \lz\ and 0.94 eV in \sa. For further discussion, we refer to these three excitons collectively as exciton \textit{A}, meaning that exciton \textit{A} in both materials consists of three degenerate BSE states. One can relate these triply degenerate excitonic states to the degeneracy of the last three VBs at $\Gamma$-point (see Fig.~\hyperref[fig:disp]{\ref{fig:disp}}). This implies that in the formation of the lowest-energy exciton, three holes are associated with the top three VBs, while the electron is purely associated with the lowest CB at the $\Gamma$ point. This type of band topology is commonly observed in III-V semiconductors \cite{bechstedt2016many}. It is important to note that the exciton \textit{A} exhibits stronger \os\ than other excitonic states in both materials. This can be explained by the orbital characters, where the strong $p$- and $s$-character at the VBM and CBM, respectively, gives rise to strong electric-dipole-allowed interband optical transitions. Hence, the largest probability for this first excitation is attributed to these transitions. Furthermore, as observed in Fig.~\hyperref[fig:ex-li]{\ref{fig:ex-li}(b)} and Fig.~\hyperref[fig:ex-sc]{\ref{fig:ex-sc}(b)}, exciton \textit{A} lies below the \gw\ band gap value of $\sim$1.5 eV for \lz\ and $\sim$1.0 eV for \sa. The eigenvalues E$^{\lambda}$ of excitons are generally associated with the binding energy E$_{b}$ of the $\lambda$th \eh\ pair, which is commonly defined as the difference between E$^{\lambda}$ and the QP-corrected band gap value. According to this definition, exciton \textit{A} has a E$_{b}$ value of $\sim$45 meV in \lz\ and $\sim$56 meV in \sa. These low values indicate the presence of weakly bound \eh\ pairs due to strong screening environment near the band gap edge. The obtained values are relatively larger in \sa\ compared to \lz, which can be attributed to the inverse proportionality of the dielectric constant with E$_{b}$ \cite{ashcroft1976}. Also, these values are nearly double the room temperature thermal energy ($\sim$25 meV). Thus, one would expect to observe bright exciton even at room-temperature, which can contribute to important features of room-temperature absorption in both HH alloys. 

In bulk semiconductors, E$_{b}$ is generally low, typically on the order of few meV, due to the presence of strong Coulombic screening effects on \eh\ pairs. Our obtained E$_{b}$ values are also within the range of other semiconductors computed in Refs. \cite{alvertis2023importance,solet2024}. Due to their relatively low E$_{b}$ values, the bound excitons in both HH materials are a type of Mott-Wannier excitons (or free excitons). This classification is supported by the fact that, for example, in the E$_{1}$ peak, the main contributions clearly coming from the VBM and CBM near the $\Gamma$ point [see Fig.~\hyperref[fig:ex-li]{\ref{fig:ex-li}(c)} and Fig.~\hyperref[fig:ex-sc]{\ref{fig:ex-sc}(c)}]. This means that exciton \textit{A} in this peak is strongly localized at this point. Consequently, one would expect delocalization behavior of exciton \textit{A} in real space, which is a direct signature of a free or Wannier-type exciton \cite{ashcroft1976,fox2010optical}. The type of excitons found in \lz\ is similar to those in GaAs, a closely related semiconductor in terms of band structure and band gap value \cite{fox2010optical}. A broader analysis of the real-space visualization of the exciton will be discussed in Sec.~\ref{sec:3c}. Furthermore, a relatively large number of weak excitations with much smaller \os s are found in the ultraviolet energy region for \sa, indicating the less excitonic effects in this region. 

To understand the relevant transitions at the marked peaks in the \iem\ spectrum associated with \eh\ pair formation, it is essential to analyze the distributions (spread) of interacting \eh\ pair (exciton) wavefunctions in both momentum and direct spaces. The hole distribution in the VB of a given excitation state $\ket{\lambda}$ in momentum space is obtained via Eq.~\hyperref[eq:wave-reci]{\eqref{eq:wave-reci-a}}, while Eq.~\hyperref[eq:wave-reci]{\eqref{eq:wave-reci-b}} estimates the electron wave function distribution for exciton $\lambda$ in the CB. The resulting plot of exciton weights, superimposed onto the DFT band structure for excitons contributing to the intense peaks labeled E$_{1}$ to E$_{4}$ in \iem\ is shown in Fig.~\hyperref[fig:ex-li]{\ref{fig:ex-li}(c)-\ref{fig:ex-li}(f)} for \lz\ and in Fig.~\hyperref[fig:ex-sc]{\ref{fig:ex-sc}(c)-\ref{fig:ex-sc}(f)} for \sa. In these figures, larger blue and red circle radii indicate a more significant contribution of the exciton to the corresponding BSE eigenstate. It is observed in \lz\ that excitons related to all four peaks are greatly associated to the last two filled VBs and the first empty CBs. However, in \sa, the excitonic features are associated to only top three VBs and first three CBs around the $\varepsilon_{\mathrm{F}}$. 

As observed in Fig.~\hyperref[fig:ex-li]{\ref{fig:ex-li}(b)} and Fig.~\hyperref[fig:ex-sc]{\ref{fig:ex-sc}(b)} for both materials, the most important contribution to the main peak (at E$_{1}$) comes from interband transitions from VBM to CBM around $\Gamma$ point in reciprocal space. Therefore, the excitonic behavior at the shoulder of the onset of the spectrum is the most critical factor for the efficient solar devices. The E$_{1}$ peak at 1.5 eV for \lz\ [see Fig.~\hyperref[fig:ex-li]{\ref{fig:ex-li}(c)}] and 0.94 eV for \sa\ [see Fig.~\hyperref[fig:ex-sc]{\ref{fig:ex-sc}(c)}] arises due to bright exciton \textit{A} around the VBM and CBM. In \lz, it has a mixed character of Zn $s$, $d$ and As $s$, $p$ states, while in \sa, it originates from the $s$ state of all three atoms along with $p$ state of C. The next peak, E$_{2}$ [Fig.~\hyperref[fig:ex-li]{\ref{fig:ex-li}(d)}], at 2.83 eV for \lz\ in visible energy region stems from the exciton character of two last filled VB and first empty CB along L-$\Gamma$-X direction, with additional minor contributions from Li $s$, $p$ and As $d$ orbitals in the vicinity of the CBM. Similarly, in \sa, excitons between the two highest occupied bands and unoccupied states in the 2–3 eV region give rise to the largest peak (E$_{2}$) in the visible spectrum, primarily driven by the $p$ states of all atoms in the occupied region, along with a significant Sc $d$ contribution in the empty region. For the E$_{3}$ and E$_{4}$ peaks in \lz, the most prominent excitonic features are found around the L point and the $\Gamma$-X path, respectively. These excitons involve all three orbitals of Zn and As atoms, with an additional contribution from the Li $s$ orbital. In \sa, excitons associated to only two main peaks in the visible light region (E$_{3}$ and E$_{4}$) are well spread over the last two occupied bands and the region extending from the CBM up to 3 eV.

\begin{figure*}[ht]
\includegraphics[width=13.7cm, height=7.4cm]{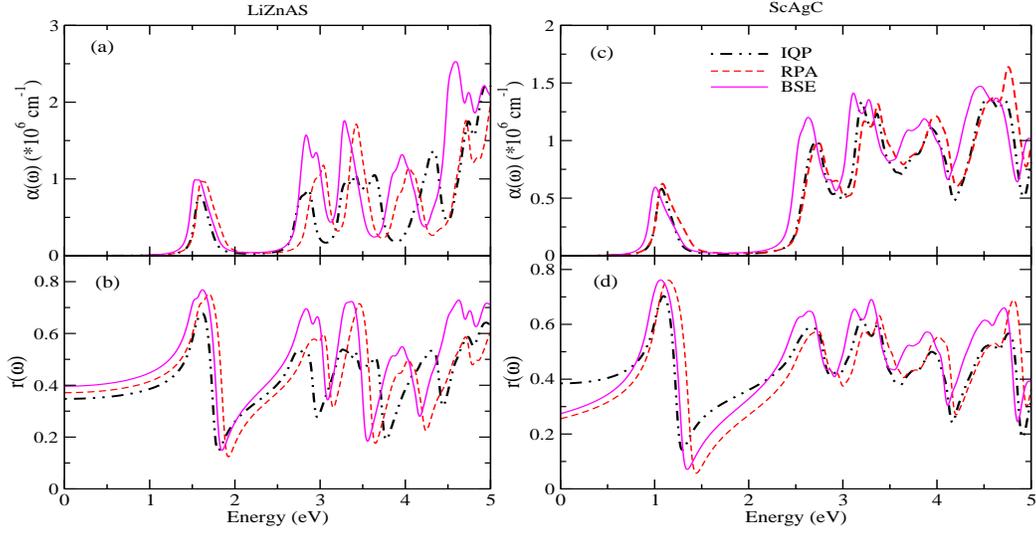} 
\caption{Absorption coefficients $\alpha (\omega)$ in (a, c) and reflectivity r($\omega$) in (b, d) obtained using methods based on IQP, RPA and BSE for both \lz\ and \sa\ compounds.} 
\label{fig:abs}
\end{figure*}

Now, moving to the next solar cell properties, Fig.~\hyperref[fig:refra]{\ref{fig:refra}} presents the results for the real n($\omega$) and imaginary k($\omega$) parts of the refractive index, obtained using the IQP, RPA, and BSE methods. Similar to the dielectric function, the strongest transition in both plots occurs at the \gw\ gap for both considered materials. In the case of \lz\ material, in both parts, compared to the IQP approximation (where neither LFEs nor \eh\ interactions are included) the intense peak is significantly affected when only LFEs are included in the RPA calculation. Furthermore, a slight redshift with an increased peak magnitude is observed when the excitonic effect is incorporated within BSE. However, both LFEs and excitonic effects do not appear to significantly affect the main peak in \sa, though a considerable excitonic effect is observed in the visible spectral range. The static value of n($\omega$) is found to range from $\sim$3.9 to 4.4 for \lz\ and from $\sim$3.2 to 4.25 for \sa. The obtained values are higher than those of other semiconducting compounds with a similar band gap range \cite{naccarato2019searching}, indicating that both studied high-refractive-index HH materials are particularly suitable for high-performance solar cell applications. Furthermore, the information on k($\omega$) provides insight into how strongly a material absorbs photons at a given $\omega$. The large peaks of k($\omega$) at the gap edge and in the visible light region indicate that the respective photons lead to a shorter penetration depth, which can be beneficial for thin-film solar cells. Beyond the main peak in the k($\omega$), comparatively stronger LFEs as well as pronounced excitonic effects are observed in \lz\ compared to \sa, where no significant differences are found between the IQP and RPA calculations of k($\omega$) in \sa.

We now compare the results of the absorption coefficient $\alpha(\omega)$ and reflectively r($\omega$) obtained from all three studied methods in Fig.~\hyperref[fig:abs]{\ref{fig:abs}}. In both materials, the $\alpha(\omega)$ starts just below the direct band gap and reaches its first main peak just after the band gap energy, as can be seen in Fig.~\hyperref[fig:abs]{\ref{fig:abs}(a)} and Fig.~\hyperref[fig:abs]{\ref{fig:abs}(c)}. This peak intensity is almost the same in the RPA and BSE methods but higher than main peak obtained from the IQP in \lz, meanwhile all three have almost the similar value for this main peak in \sa. However, the peak position is slightly shifted toward lower energy in BSE compared to the other two. In the visible light region, a strong effect of both LFEs and \eh\ interactions is observed around an energy of 2.8 eV in \lz, while only the latter effect significantly influences the spectrum around 2.6 eV in \sa. The values at these energies are estimated to be $\sim$1.6 and 1.2 $\times$ $10^6$ cm$^{-1}$ for respective compounds. Beyond the visible region, both HH compounds exhibit considerable LFEs and excitonic effects, but their contribution is less significant in \sa. The high optical absorption capabilities in the visible to ultraviolet spectral regions make both materials highly suitable for visible-light solar harvesting applications as well as for use in UV sensing devices. Similarly, a significant difference between calculated r($\omega$) from the IQP, RPA and BSE methods is noticeable in Fig.~\hyperref[fig:abs]{\ref{fig:abs}(b)} and Fig.~\hyperref[fig:abs]{\ref{fig:abs}(d)}. The main peak around the band gap value appears with greater magnitude in both the RPA and BSE spectra than in the IQP. The lowest value in the visible spectrum is found to be below 40\% in both materials. However the minimum r($\omega$) value within the studied energy window occurs at an energy beyond the band gap value in both materials. This value from BSE is almost 16\% at $\sim$1.8 eV in \lz, while it is $\sim$7\% at $\sim$1.3 eV in \sa. Such low values indicate that these materials could be used for antireflection coating surfaces.
\begin{figure*}[ht]
\includegraphics[width=15.5cm, height=9.7cm]{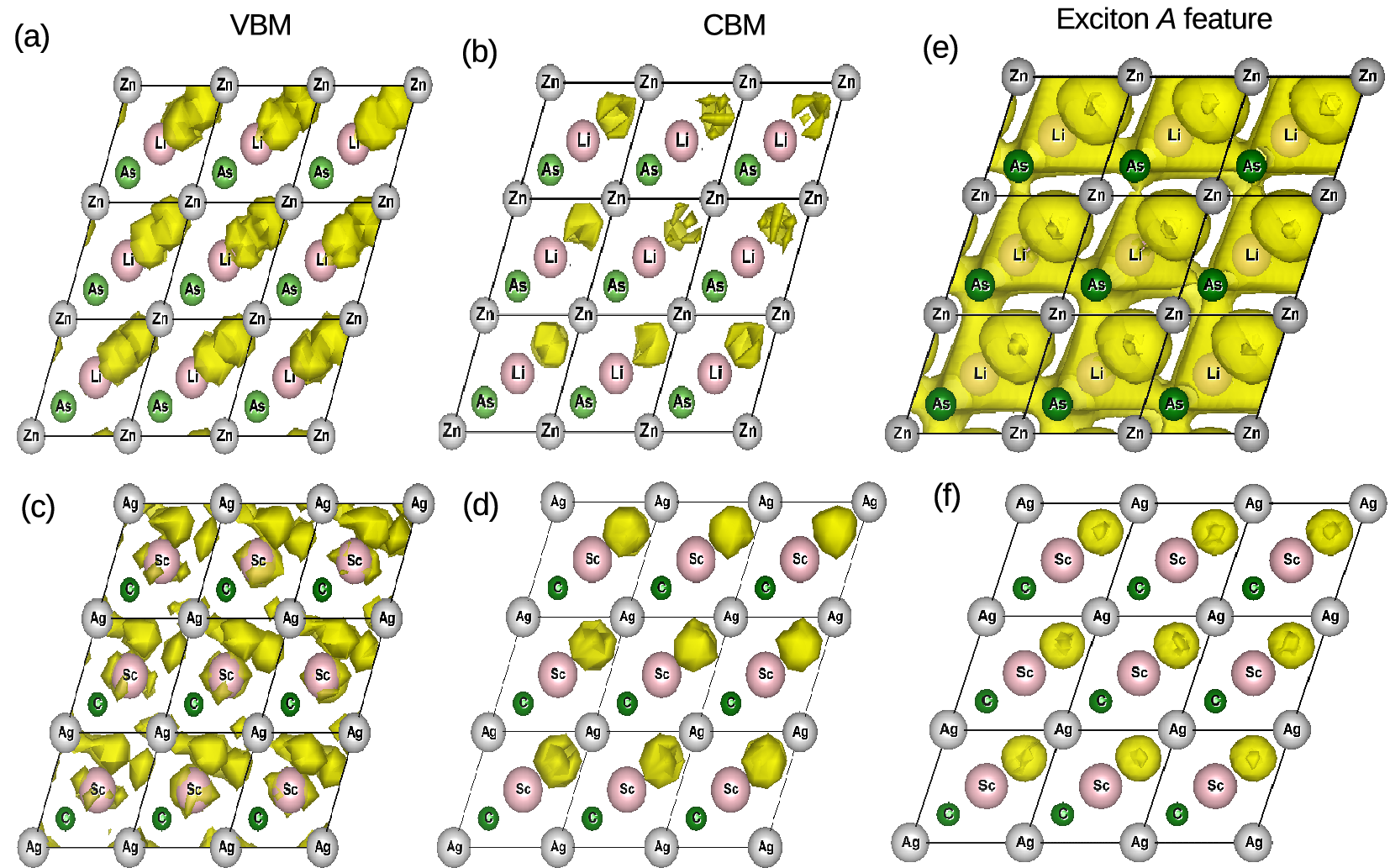} 
\caption{Real-space representation of the charge density distribution at the valence band maxima (VBM) for \lz\ [in (a)] and \sa\ [in (c)], and at the conduction band minima (CBM) for \lz\ [in (b)] and \sa\ [in (d)], calculated using DFT. The spatial extent of the electronic part of the exciton \textit{A} wave function $\Psi_{\lambda}(\textbf{r}_{e}, \textbf{r}_{h})$ related to the E$_{1}$ peak in the \iem\ spectrum is shown for \lz\ [in (e)] and \sa\ [in (f)]. The hole position $\textbf{r}_{h}$ is fixed near the lithium and scandium atoms at (0.52, 0.52, 0.52). A 3 $\times$ 3 $\times$ 3 supercell is used, and all plots are viewed along the \textbf{c} axis with an isosurface level of 6.2 $\times$ 10$^{-6}$, as visualized using the VESTA software. } 
\label{fig:ex1-rs}
\end{figure*}

\subsection{\label{sec:3c}Real-space projection of excitons}

The localized (or delocalized) nature of the hole and electron amplitudes in reciprocal space is highly correlated with the spread of the excitonic wave function in real space. Therefore, we discuss the real-space representation of the excited electron wave function for an exciton contributing to the intense peaks in absorption spectrum. As seen in Eq.~\hyperref[eq:wave-real]{\eqref{eq:wave-real}}, the two particle \eh\ wave function is a six-dimensional object (three coordinates for electron and three for hole). Therefore, it is convenient to fix the position of one of the particles and visualize exciton wave function with respect to the coordinates of the other. For our analysis, we fix the hole position slightly away from the lithium site in \lz\ and the scandium site in \sa\ at (0.52, 0.52, 0.52). We plot these real space probabilities using VESTA plotting software \cite{momma2011vesta} and set an isosurface value of 6.2 $\times$ 10$^{-6}$. 

Since exciton \textit{A} has the largest \os\ among the excitons in the studied energy region, we focus our analysis on its distribution in real space. The isosurface plot of the electronic part of this exciton wave function relative to the hole is shown in the lattice structure of both HH compounds in Fig.~\hyperref[fig:ex1-rs]{\ref{fig:ex1-rs}(e-f)}. Since the electron and hole amplitudes of this exciton are highly concentrated within a small region in \textbf{k}-space (near the $\Gamma$ point), this feature can be better visualized after analyzing the real space probability of finding the electron at the CBM and the hole at the VBM when the \eh\ interaction is completely absent, \textit{i.e.}, at the DFT level. These charge density distribution into DFT picture in real space at the band extrema (VBM and CBM) for both compounds are plotted in Fig.~\hyperref[fig:ex1-rs]{\ref{fig:ex1-rs}(a-d)}. These electron/hole distribution features in Fig.~\hyperref[fig:ex1-rs]{\ref{fig:ex1-rs}}, for both the non-interacting and interacting \eh\ cases, are visualized using a 3 $\times$ 3 $\times$ 3 supercell in a 3D view along the \textbf{c} axis. In this figure, the isosurface provides insight into the overall spread of the hole at the VBM (left panels), the electron at the CBM (middle panels), and exciton \textit{A} around the band gap (right panels).

In Fig.~\hyperref[fig:ex1-rs]{\ref{fig:ex1-rs}(a-d)}, for both materials, the hole density at the VBM is more widely spread in a single (primitive) unit cell compared to the electron density at the CBM. In \lz, the probability of finding the electron/hole is higher at slightly away from the center in a particular direction towards Zn in each unit cell. This indicates that the electron/hole distribution in this supercell is not localized at any of the atoms in \lz. However, for \sa, the hole wave function is distributed almost throughout the primitive unit cell, except in the immediate vicinity of the carbon atom, which may be due to the fact that the reciprocal space projection shows the main contribution of the C $p$ states at the VBM [see Fig.~\hyperref[fig:bandch-sc]{\ref{fig:bandch-sc}}]. Furthermore, the hole density is observed to be maximum around the Sc atom and weaker around Ag, indicating that it is localized at Sc atom in all primitive unit cells. However, the electron probability distribution exhibits the opposite behavior, which has maximum intensity just away from Sc, towards Ag [see Fig.~\hyperref[fig:ex1-rs]{\ref{fig:ex1-rs}(d)}]. Finally, it is important to note in both materials that the spread of the wave function of the hole/electron is observed across several primitive cells, which means both particles are very delocalized in real space.  

Figure~\hyperref[fig:ex1-rs]{\ref{fig:ex1-rs}(e-f)} illustrates the real-space distribution of excited electron of exciton \textit{A} when \eh\ interaction is taken into account. The electron distribution for \lz\ in Fig.~\hyperref[fig:ex1-rs]{\ref{fig:ex1-rs}(e)} clearly demonstrates how the physical picture can be counterintuitive if one relies only on the single-particle DFT framework. It is worth highlighting the strong localization of this exciton at the Li atom, meaning that the electron probability distribution is significantly affected compared to the DFT results. This is a clear signature of the strong excitonic effect at the gap edge in \lz, which was previously observed in the \iem\ spectrum. For \sa\ as shown in Fig.~\hyperref[fig:ex1-rs]{\ref{fig:ex1-rs}(f)}, no significant change is observed in the distribution nature over the supercell compared to the electron density distribution at the DFT level. The exciton \textit{A} wave function can then be approximately expressed as the product of the hole wave function at VBM and electron wave function at CBM due to nearly similar charge density distributions for VBM/CBM and exciton \textit{A} wave function. Due to this behavior, the role of excitonic states A$^{\lambda}$ is weaker, suggesting a weak excitonic effect in the main absorption peak of \sa. Furthermore, in both materials, exciton \textit{A} starts to spread beyond the primitive unit cell, further illustrating our observation of the Mott-Wannier character discussed in Sec.~\ref{sec:3b}. An important factor for PV applications is that excitonic absorption does not directly produce mobile charge carriers. The exciton must first dissociate, but there is a possibility that carriers may recombine before contributing to the current. If excitons are delocalized in space, the likelihood of recombination may decrease because free charge carriers resulting from exciton dissociation are primarily present in the neutral region (where recombination probability is comparatively lesser than in other regions) of \emph{pn} junction solar cells \cite{nelson2003the}. Therefore, the presence of Mott-Wannier excitons in these materials is a positive sign for solar cell applications. 

It is worth noting that the present BSE calculations account only for dielectric screening effects on \eh\ interacting pairs due to other electrons, meaning that W in Eq.~\hyperref[eq:W]{\eqref{eq:W}} includes only the electronic contributions to screening. Consequently, quantities such as dielectric constant and E$_{b}$ are obtained while neglecting additional screening contributions, such as those arising from phonons \cite{filip2021phonon,lee2024phonon,umari2018infrared}. These quantities can be modified if the phonon screening effects has to be considered in the BSE Kernel in Eq.~\hyperref[eq:L]{\eqref{eq:L}} \cite{alvertis2024phonon,schebek2024phonon}. Such effects have been observed in recent studies \cite{filip2021phonon,lee2024phonon,adamska2021bethe}, where the inclusion of phonon screening leads to modified E$_{b}$ values in various types of semiconductors that show better agreement with experimental results. For example, the phonon screening reduces the ion-clamped E$_{b}$ values for CsPbX$_3$ (X = Cl, Br, I) perovskites by almost 12-17\% \cite{filip2021phonon}, and a reduction of almost 30\% is observed in bulk GaN \cite{lee2024phonon}. This indicates that an enhanced screening environment due to phonons leads to a reduction in exciton binding energies. Accordingly, such a reduction can also be expected in our calculated values of $\sim$45 meV (for \lz) and $\sim$56 meV (for \sa) when phonon screening is taken into account. The decrease in E$_{b}$ values is a favorable attribute of these HH materials for PV applications. A detailed discussion of lattice screening effects on exciton behavior in these HH compounds is deferred to a future study, as it could enhance understanding of their suitability for PV applications.

\subsection{\label{sec:3d}Solar cell efficiency}

Now to evaluate the photoinduced carrier conversion efficiency of a single-junction solar cell device, we have used a theoretical efficiency model based on current-voltage analysis, known as the spectroscopic limited maximum efficiency (SLME) model \cite{yu2012identification}. The SLME can be determined from the maximum power output density (P$_{\mathrm{max}}$) when sunlight with intensity (I$_{\mathrm{sun}}$) is incident on the solar cell \cite{yu2012identification,solet2024},
\begin{eqnarray} \label{eq:slme}
\mathrm{SLME} = \frac{P_{\mathrm{max}}}{\int^{\infty}_{0}\varepsilon I_{\mathrm{sun}}(\varepsilon)d\varepsilon} .
\end{eqnarray}  
where P$_{\mathrm{max}}$ is obtained by numerically maximizing the product of the total current density (J) and voltage (V) over the absorbing material. I$_{\mathrm{sun}}$ at photon energy $\varepsilon$ is taken to be the standard AM1.5G spectrum at 25 $^\circ\text{C}$, provided by the National Renewable Energy Laboratory \cite{AM1.5}. Furthermore, J of a \textit{pn}-junction solar cell under light at temperature T is given by, 
\begin{eqnarray}\label{eq:j}
J = J_{\mathrm{sc}} - J_{0}(e^{eV/k_{B}T} - 1)
\end{eqnarray} 
The term $J_{\mathrm{sc}}$ is the short-circuit current density due to incident photons and can be defined as, 
\begin{eqnarray}\label{eq:jsc}
J_{\mathrm{sc}} = e\int^{\infty}_{0} (1 - e^{-2\alpha(\varepsilon)L}) I_{\mathrm{sun}}(\varepsilon) d\varepsilon
\end{eqnarray}
where $e$ is the elementary charge, $\alpha(\varepsilon)$ is the absorption coefficient, and $L$ is the thickness of the layer, assuming zero reflectivity from the front surface and unity reflectivity from the back surface (hence the factor of 2).

In Eq.~\hyperref[eq:j]{\eqref{eq:j}}, J$_{0}$ represents the current density resulting from both radiative and nonradiative \eh\ recombination processes. In the SLME formalism \cite{yu2012identification}, J$_{0}$ has a simplified form as J$^{\mathrm{r}}_{0}$/f$_{\mathrm{r}}$, where f$_{\mathrm{r}}$ is the fraction of the radiative \eh\ recombination current (J$^{\mathrm{r}}_{0}$), approximated by $e^{-\Delta/k_{B}T}$. Here, $\Delta$ denotes the energy difference between the direct-allowed band gap and the minimum band gap of the material. Finally, J$^{r}_{0}$ is derived using the detailed balance limit in the presence of black-body photon flux (I$_{\mathrm{bb}}$) at temperature T as,
\begin{eqnarray}\label{eq:jr}
J^{r}_{0} = e\pi\int^{\infty}_{0} (1 - e^{-2\alpha(\varepsilon)L}) I_{\mathrm{bb}}(\varepsilon, T) d\varepsilon
\end{eqnarray}

Therefore, the spectroscopic quantities such as band gap and absorption coefficient are essential to determine the thickness-dependent SLME of a thin-film solar cell. SLME is an improved version of the Shockley and Queisser (SQ) efficiency, incorporating the effects of the radiative recombination fraction f$_{\mathrm{r}}$ and the photon absorptivity $a(\varepsilon)$ = 1 - e$^{-2\alpha(\varepsilon)L}$ \cite{yu2012identification}. SQ efficiency assumes f$_{\mathrm{r}}$ = 1, implying that all \eh\ recombination losses are purely radiative. Additionally, it models photon absorptivity as a step function \cite{yu2012identification}. 

\begin{figure}[ht]
\includegraphics[width=7.1cm, height=6.3cm]{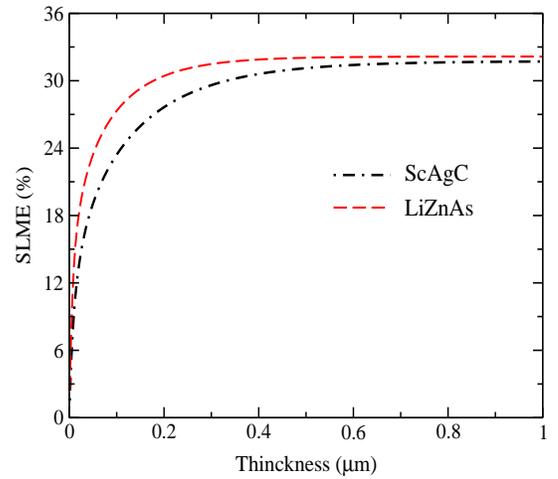} 
\caption{Calculated room temperature spectroscopic limited maximum efficiency (SLME) as a function of thickness of both materials using absorption coefficient from Bethe-Salpeter equation method.} 
\label{fig:slme}
\end{figure}

Figure~\hyperref[fig:slme]{\ref{fig:slme}} shows the results of the room-temperature SLME with respect to the film thickness of the considered solar materials. The BSE-based $\alpha(\omega)$ is used in the estimation. The figure shows that SLME increases rapidly at lower thicknesses up to 0.1 $\mu$m, then increases very slowly and becomes nearly constant after almost 0.4 $\mu$m. The highest values for thin-film thickness at $\sim$0.4 $\mu$m are obtained to be $\sim$32\% for \lz\ and $\sim$31\% for \sa. The SLME for \sa\ is lower than that of \lz\ at lower thicknesses, while both attain nearly similar values ($\sim$32.1\% for \lz\ and 31.7\% for \sa) at thicker layer ($\sim$1 $\mu$m). These values closely approach the SQ efficiency limits of almost 32.15\% for \lz\ and 31.78\% for \sa. This indicates that the SLME approaches the SQ limit at higher thicknesses due to the photon absorptivity approaching unity in thicker layers, as well as f$_{\mathrm{r}}$ being equal to 1 as a result of the direct band gaps. The SLME value at 0.4 $\mu$m for \lz\ is significantly higher (more than double) than that of GaAs, which was reported to be almost 15\% at the same thickness \cite{yin2015superior}. However, the highest predicted SLME for GaAs is $\sim$28\% at $\sim$3 $\mu$m, which is still lower than the $\sim$32.15\% for \lz. Therefore, \lz\ can be considered one of the best alternatives to GaAs-based thin film solar cells. The obtained SLME values are also higher than those of many other HH compounds \cite{sahni2020reliable,zhang2012sorting} and are comparable to that of NaZnSb (31.17\%) \cite{sahni2020reliable}. These high efficiencies indicate that both HH materials are highly suitable for thin-film single-junction PV solar cell devices.

It is important to note that the SLME obtained here is calculated by considering only radiative \eh\ recombination inside a solar cell. Since the radiative process is a direct band-to-band recombination, it does not require any external agency to complete this process \cite{nelson2003the}. This means it is a single-step process occurring within the solar material. However, in semiconducting PV cells, various types of \eh\ recombination can occur, such as non-radiative recombination \cite{nelson2003the}. This is a two-step process in which defect states facilitate the recombination. Due to its multi-step process, the current contribution from \eh\ pair losses via non-radiative decay is lower compared to the one-step radiative process when evaluating the total current density of illuminated solar cells under sunlight at a given bias voltage. This behavior is observed in many direct and indirect band gap semiconducting materials \cite{solet2024,moustafa2024selenium}. For instance, in the direct band gap Ca$_2$Si, the highest predicted SLME was 31.2\%, which decreased to be 28.5\% when non-radiative \eh\ recombination was included \cite{solet2024}. In indirect gap materials such as Mg$_2$Si, the maximum SLME of 1.3\% was reduced to be 1.2\% due to non-radiative effects \cite{solet2024}. On the other hand, the studied defects in indirect selenium semiconductor have been found to exhibit zero non-radiative recombination \cite{moustafa2024selenium}. Based on these observations in various materials, one can conclude that in the presence of non-radiative recombination, the maximum solar efficiency for both \lz\ and \sa\ materials is expected to be around 29-30\%. Since \lz\ has already been synthesized, we strongly recommend synthesizing \sa\ as well. Finally, we believe that the experimental community should explore these HH materials for the innovative development of highly efficient single-junction solar PV devices.

\section{Conclusions} \label{sec:concl}

In this work, we perform an \emph{ab initio} study of two half-Heusler compounds, \lz\ and \sa, for solar cell applications by obtaining their electronic and optical response properties from ground-state DFT to excited-state many-body methods. The fundamental band gap, which is direct in nature, is estimated to be $\sim$0.6 (1.5) eV and $\sim$0.45 (1.0) eV using KS-PBE (\gw) calculation for \lz\ and \sa, respectively. We estimate the optical properties at three different levels of theory$-$ IQP, RPA, and excitonic effects within the BSE formalism. The largest peak of \iem\ is observed at the direct optical band gap, with values of $\sim$52 (87), 77 (87), 88 (91) from the respective methods for \lz\ (\sa). Orbital character analysis shows that this main peak originates primarily from $p\rightarrow s$ interband transitions. In both materials, we find triply degenerate, loosely bound bright excitons (exciton \textit{A}), below the direct optical gap, with binding energies in the range of $\sim$45-56 meV. The strongest \os\ of excitons is observed at the band gap, while relatively lower strengths are observed in the visible and ultraviolet energy regions. However, \lz\ exhibits large \os s compared to \sa. We further obtain the exciton amplitude related to intense optical interband transitions in the reciprocal space. Exciton \textit{A} is found to be highly localized around the band gap region in reciprocal space, while it is highly delocalized in real space. We finally obtain the thickness-dependent solar efficiency using SLME calculations. The highest obtained SLME values are $\sim$32\% for \lz\ and $\sim$31\% for \sa\ at $\sim$0.4 $\mu$m layer thickness. These results highlight the important role of excitonic states in the solar energy absorption process and provide valuable information for developing highly efficient next-generation single-junction thin-film solar cells.

\vspace{0.2in}
\section*{Acknowledgement}
We acknowledge the computational support provided by the High-Performance Computing (HPC) PARAM Himalaya at the Indian Institute of Technology Mandi.

\bibliography{paper}

\end{document}